\documentclass[%
superscriptaddress,
 amsmath,amssymb,
 aps,
 pra,
 twocolumn 
 ]{revtex4-2}

\usepackage{graphicx}
\usepackage{dcolumn}
\usepackage{bm}
\usepackage{xcolor}
\usepackage[caption=false]{subfig}
\usepackage{xfrac}
\usepackage{blindtext}
\usepackage{times}
\usepackage{comment}
\usepackage{simpler-wick}
\usepackage{hyperref}
\usepackage{chngpage}

\newcommand{\fading}[1]{{\color[cmyk]{0.0,1.0,1.0,0.0}#1}}

\definecolor{darkblue}{HTML}{004D6B}
\definecolor{darkred}{HTML}{8c1515}
\hypersetup{
	colorlinks=true,
	urlcolor=darkblue,
	citecolor=darkred,
	linkcolor=darkblue,
	breaklinks
}

\newcommand\scalemath[2]{\scalebox{#1}{\mbox{\ensuremath{\displaystyle #2}}}}

\begin{document}


\title{Where are the photons in a transmission-line pulse?}

\author{Evangelos Varvelis}
\affiliation{Institute for Quantum Information, RWTH Aachen University, 52056 Aachen, Germany}
\affiliation{J\"ulich-Aachen Research Alliance (JARA), Fundamentals of Future Information Technologies, 52425 J\"ulich, Germany}

\author{Debjyoti Biswas}
\affiliation{Department of Physics, IIT Madras, Chennai 600036, India}

\author{David P. DiVincenzo}

\affiliation{Institute for Quantum Information, RWTH Aachen University, 52056 Aachen, Germany}
\affiliation{J\"ulich-Aachen Research Alliance (JARA), Fundamentals of Future Information Technologies, 52425 J\"ulich, Germany}
\affiliation{Peter Gr\"unberg Institute, Theoretical Nanoelectronics, Forschungszentrum J\"ulich, 52425 J\"ulich, Germany}%


\date{\today}

\begin{abstract}
We develop a photonic description of short, one-dimensional electromagnetic pulses, specifically in the language of electrical transmission lines. Current practice in quantum technology, using arbitrary waveform generators, can readily produce very short, few-cycle pulses in microwave TEM guided structures (coaxial cables or coplanar waveguides) in a very low noise, low temperature setting. We argue that these systems attain the limit of producing pure coherent quantum states, in which the vacuum has been displaced for a short time, and therefore short spatial extent. When the pulse is bipolar, that is, the integrated voltage of the pulse is zero, then the state can be described by the finite displacement of a single mode. Therefore there is a definite mean number of photons, but which have neither a well defined frequency nor position. Due to the Paley-Wiener theorem, the two-component photon ``wavefunction" of this mode is not strictly bounded in space even if the vacuum displacement that defines it is bounded. This wavefunction's components are, for the case of pulses moving in a specific direction, complex valued, with the real and imaginary parts related by a Hilbert transform. They are thus akin to the ``analytic signals" of communication theory. When the pulse is unipolar no photonic description is possible -- the photon number can be considered to be divergent. We consider properties that photon  counters and quantum non-demolition detectors must have to optimally convert and detect the photons in several example pulses, and we discuss some consequence of this optimization for the application of very short pulses in quantum cryptography.   
\end{abstract}

\maketitle

\section{Introduction}

The arbitrary waveform generator (AWG) \cite{aboutAWG} is a key part of the instrumentation of many present-day quantum-technology devices. In a solid-state quantum computer, it is used, together with other microwave components like RF signal generators, to deliver controlling radiation, via transmission lines, to the immediate vicinity of the qubits. Appropriately pulsed signals vary the contributions in the computer's Hamiltonian, or cause quantum measurements to be performed. From this point of view, they are part of the classical apparatus that manipulates the quantum world of the quantum computer.

But we can alter our point of view and ask, what quantum states describe the signals that the AWG can produce? We will in particular address the question, what is the nature of the photons present in an AWG signal? We will focus on pulsed signals with a definite starting and stopping time, so that we expect that photons will appear in a burst. But what is the starting and ending time of this burst of photons? How many photons are there, and what are their attributes (e.g., frequency)? This paper will give a definite prescription for calculating these properties.

While the AWG can emit a strictly localized pulse in the sense of its voltage profile $V(x)$, we confirm that despite this, the photons this pulse contains cannot be deemed to be strictly localized. This effect has been long discussed in field theory \cite{landau1930quantum,RevModPhys.21.400}. Recent work has begun to explore the surprising properties of this dual manifestation of 1D electromagnetic pulses \cite{virally2019unidimensional}. Previous work on the ``wavefunction of the photon" \cite{bialynicki1994wave} establishes that some degree of localization is possible \cite{Saari06}, but is constrained by the Paley-Wiener theorem \cite{PaleyWiener}, from which one can conclude that the photon wavefunction must be nonzero over all space. This can be seen explicitly in our calculations, where these wavefunctions must have power-law tails. This is by contrast to the case of photons in 3D, where exponential localization is possible  \cite{bialynicki1998exponential}.

We will also calculate the relationship between the voltage pulse $V(x)$ and the mean total number of photons $\langle n\rangle$, with some interesting implications for quantum communication protocols. If $V(x)$ is unipolar, i.e., has a nonzero integrated value, then $\langle n\rangle$ is undefined: due to an infrared divergence, the pulse can be viewed as having an unbounded photon number. Thus, such a signal is never appropriate for quantum cryptography \cite{RevModPhys.92.025002}: no matter how small $V(x)$ is, the pulse is susceptible, if Eve has an optimized set of instruments, to a splitting attack followed by re-amplification. For a bipolar pulse ($\int V(x)dx=0$), $\langle n\rangle$ is finite, but its dependence on the pulse shape is nontrivial: we show an example of a split pulse (two parts of a pulse separated by distance $w$) where naive arguments based on mean frequency would estimate a photon number independent of $w$, while in fact $\langle n\rangle$ goes like $\log{w}$. Thus, an Alice and Bob with only knowledge of standard frequency-selective detectors may conclude that a pulse is dim, $\langle n\rangle<1$, while the pulse may be, for Eve with an optimal detector, quite bright and easily attackable. The final part of our paper will discuss the attributes that an optimal photon detector must have in the pulsed setting.  


\section{Transmission Line Basics}

The AWG is set up to deliver an arbitrary voltage function $V_{out}(t)$ at the output terminals. {\em Arbitrary} means that the voltage, while a continuous function of time, has a different, arbitrarily chosen value every 250ps or so (some AWGs have faster ``sampling rates").
Thus, its frequency content will be in the microwave band (or lower). While the AWG signal is often mixed with that from an ac signal generator to modulate a tone of a definite frequency, it can be, and in some cases is, simply launched into a transmission line. 

The fundamental parameters of this transmission line, which will be relevant for subsequent analysis, can be taken to be the wave impedance $Z_0$ and the velocity $v$. We will also make reference to an alternative pair of parameters $\ell$, the inductance of the line per unit length, and $c$, the capacitance per unit length. These parameters are interrelated by the formulas $Z_0=\sqrt{\ell/c}$, $v=1/\sqrt{\ell c}$.

Classically, the transmission line transmits waveforms of any shape with velocity $+v$ or $-v$. Of course, the output of the AWG moves in one direction only on a perfect transmission line (let us call it right moving), so it contributes a time-evolved voltage $V(x,t)=V_{out}(t-x/v)$ for $x\ge 0$. The transmission line also carries a current $I(x,t)$, which in the right-moving case is just proportional to $V(x,t)$ with proportionality $1/Z_0$. But let us review the general relation which results if signals with components with both velocity $+v$ and $-v$ are present. This will occur if there are reflections due to imperfections or discontinuities in the transmission line. It is readily shown \cite{pozar2012} that if the voltage signal is given by the general expression
\begin{equation}
V(x,t)=f_R(t-x/v)+f_L(t+x/v),
\end{equation}
then the current function is
\begin{equation}
I(x,t)=\frac{1}{Z_0}\left(f_R(t-x/v)-f_L(t+x/v)\right).
\end{equation}
This means that at any instant of time, $V(x,t=0)$ and $I(x,t=0)$ can be two entirely independent functions of $x$, but with subsequent time evolution determined by the two functions
\begin{eqnarray}
    f_R(t-x/v)&=&\frac{1}{2}(V(x,0)+Z_0 I(x,0))\\
     f_L(t-x/v)&=&\frac{1}{2}(V(x,0)-Z_0 I(x,0))\label{rightmove}
\end{eqnarray}
For the signal as emitted by the AWG, $f_L=0$.

\section{Quantum-State Description of a Pulse Emitted by an AWG}

We wish to interpret the classical field quantities just discussed as expectation values of particular quantum field operators in a quantum state. We will pursue this using the quantization procedure employed in circuit quantum electrodynamics (cQED). Following the scheme presented in \cite{blais2021circuit}, we take the two fields describing the transmission line circuit (cf.~Fig.~3 of \cite{blais2021circuit}) to be the flux field $\hat\Phi(x)$ and the charge-density field $\hat Q(x)$. These are Hermitian fields, with commutator $[\hat\Phi(x),\hat Q(x')]=i\hbar\delta(x-x')$, with which we express the transmission-line Hamiltonian
\begin{equation}
    H=\int_{-\infty}^\infty dx\left\{\frac{1}{2c}\hat Q(x)^2+\frac{1}{2\ell}[\partial_x\hat\Phi(x)]^2\right\}.
    \label{hamtrans}
\end{equation}

When we consider the possible quantum-state description of a given AWG signal, we must constrain the state such that it has the appropriate expectation values of these operators. We can use basic circuit relations to give the needed relation between the expectation values of our quantum fields $\hat\Phi$ and $\hat Q$ and the classical field variables $V$ and $I$. We work at a particular time, which we call zero:
\begin{eqnarray}
    \varphi(x)&\equiv&\langle\hat\Phi(x)\rangle=-\ell\int I(x,t=0)dx\label{rel1},\\
    q(x)&\equiv&\langle \hat Q(x)\rangle=c\,\, V(x,t=0).\label{rel2}
\end{eqnarray}
The first equation can equivalently be written $d\varphi(x)/dx=-\ell I(x)$, which is understood by noting that $d\varphi(x)$ is the magnetic flux produced by current $I$ flowing through inductor $\ell dx$. We present Eq.~(\ref{rel1}) as an indefinite integral, leaving for later the important discussion of the appropriate integration constant. 

With these preliminaries, we consider the question: what is the quantum state emitted by the AWG? Besides the fact that it has certain specified expectation values, we know very little about it. Being emitted by a macroscopic device, it is very likely to be a mixed state. We assume that the emission does not vary very much from shot to shot, constraining somewhat the properties of this mixed state. In particular, it should not have too large a value of the variances of the field quantities. 

We speculate no further on what this quantum state might be, but we consider further a very common use of this emitted state \cite{krinner2019engineering}: it is passed into a region of very low temperature, and it is attenuated very strongly. Ideally, the attenuator diminishes the amplitude of all (frequency modes) equally, and is reflectionless -- the textbook resistive-tee attenuator has these properties \cite{pozar2012}. Being very cold, the attenuator emits very few thermal photons into the transmission line.

Under these conditions, we can say something more definite about the likely state of an AWG pulse after attenuation. According to the standard model of pure loss (\cite{Walls2008}, Sec.~6.2.5), the Wigner function of any state whose initial amplitude in mode $\omega$ is $\langle\alpha_\omega\rangle$, subject to a large loss by factor $\lambda$ ($\lambda\ll 1$), approaches 
\begin{equation}
    W(\alpha)=\frac{2}{\pi}\exp{(-2|\alpha-\lambda\langle\alpha_\omega\rangle|^2)}.
\end{equation}
This is the Wigner function of the single-mode pure coherent state. The corrections to this will be very small so long as the standard deviation $\sigma$ of the initial state is reduced by attenuation to the half-photon level:
\begin{equation}
 \lambda\sigma_\omega\alt\frac{1}{2}.   
\end{equation}
Since 30dB attenuation is common ($\lambda=1/1000$), such conditions should be feasible to satisfy.


We proceed with the hypothesis that the pulsing of the AWG results in the creation of a multimode coherent state of the sort first introduced by Glauber \cite{glauber1963quantum} to discuss the quantum state of laser radiation. The foregoing should be only considered as a heuristic justification, rather than a rigorous proof, of this hypothesis. It is hopefully better than the ``convenient fiction" of describing laser radiation with such a state \cite{PhysRevA.55.3195}. We will be quite busy shortly in considerably sharpening the notion of the state of photons in our hypothesized AWG pulse.

A coherent state $|\Psi_0\rangle$ with the desired expectations of the field functions $q(x)$ and $\varphi(x)$ (Eqs.~(\ref{rel1},\ref{rel2})) at time $t=0$ is written as a displacement operator acting on the vacuum:
\begin{eqnarray}\label{HermDisp}
    |\Psi_0\rangle = \exp\scalemath{1.00}{\left[\frac{i}{\hbar}\int_{-\infty}^\infty \left( q(x)\hat\Phi(x)-\varphi(x)\hat Q(x)\right)dx\right]}|0\rangle.
\end{eqnarray}
We will be considering pulses for which we have set the origin of time such that they have already travelled a considerable distance past the attenuator, so that we can take the $x$ integration to $-\infty$ as indicated.

\section{Photonic Content of Transmission Line Pulse}

We now come to the central question of this paper: what are the attributes of photons making up this state? For this we take the point of view that Eq.~(\ref{HermDisp}) is a superposition of states of different photon numbers, but where the photons are those of a single mode. Said mathematically, this means that we expect to be able to rewrite Eq.~(\ref{HermDisp}) in the form \cite{virally2019unidimensional} 
\begin{equation}\label{OneModeDisp}
    |\Psi_0\rangle=
\exp(\beta b^\dagger-\beta^* b)|0\rangle.
\end{equation}
This introduces the number $\beta$ and the quantum operators $b$, $b^\dagger$. As usual, $|\beta|^2$ can be interpreted as the mean number of photons in the pulse. The operators $b$, $b^\dagger$ should have the properties $[b,b^\dagger]=1$, $b|0\rangle=0$. We note that this final property permits us to rewrite the state in Eq.~(\ref{OneModeDisp}) as $\exp(\beta b^\dagger)|0\rangle$. 
But it will be convenient to proceed by matching the full displacement operator of Eq.~(\ref{OneModeDisp}) with that of Eq.~(\ref{HermDisp}).

We will find that for pulses satisfying some conditions, which we will derive, it will be possible to make this identification. If the conditions are not satisfied, the identification will fail due to a divergence of the displacement parameter $\beta$. Note that when $\beta$ exists, its phase can always be absorbed into the phases of the operators $b$ and $b^\dagger$, and therefore wlog we will take $\beta$ to be positive and real; we will see that this makes the identification of $\beta$, $b$, and $b^\dagger$ unique.

To get started on finding these quantities, we note that operator $\beta b^\dagger$ must be a functional of $\hat\Phi(x)$ and $\hat Q(x)$, since these are the only quantum field operators in the problem. Thus we write
\begin{equation}
    \beta b^\dagger=\frac{1}{2\hbar}\int_{-\infty}^\infty \left( \theta_q(x)\hat\Phi(x)-\theta_\varphi(x)\hat Q(x)\right)dx,\label{creation}
\end{equation}
introducing the new coefficient (c-number) functions $\theta_q(x)$ and $\theta_\varphi(x)$. Note that $\beta b^\dagger$ can be written as a hermitian plus an antihermitian part, and that the antihermitian part is immediately given by the antihermitian argument of the exponential function in the displacement operator of Eq.~(\ref{HermDisp}). Therefore we have
\begin{eqnarray}
    {\mbox{\rm Im}}[\theta_q(x)]&=&q(x),\nonumber\\
    {\mbox{\rm Im}}[\theta_\varphi(x)]&=&\varphi(x).\label{HermPart}
\end{eqnarray}
Thus the work to be done is reduced to finding the hermitian 
part, that is, the real part of these functions. 

We will make use of
the eigenmode creation operators of the infinite transmission line, which have the form \cite{LSZ,Greiner}
\begin{equation}
    a_p^\dagger=\frac{1}{2\sqrt{\pi\hbar}}\int_{-\infty}^\infty dx\ e^{-ipx} \left(\sqrt{cv|p|}\hat\Phi(x)-\frac{i}{\sqrt{cv|p|}}\hat Q(x)\right).\label{EigenCreate}
\end{equation}
When running the wavevector $p$ from $-\infty$ to $\infty$ this set of operators is complete -- thus, the set $a_p$ span the algebra of the operators that annihilate the vacuum. $b^\dagger$ should thus be taken as a linear combination of the $a_p^\dagger$ (i.e., integral over $p$) \cite{glauber1963coherent,titulaer1965correlation}. We can identify this linear combination by inserting the expressions for the quantum fields $\hat\Phi(x)$ and $\hat Q(x)$ in terms of these mode creation and annihilation operators \cite{blais2021circuit}:
\begin{eqnarray}
    \hat\Phi(x)&=&\frac{1}{2}\int_{-\infty}^\infty dk \sqrt{\frac{\hbar}{\pi vc|k|}}\left(e^{ikx}a_k^\dagger+e^{-ikx}a_k\right)\! ,\label{PhiToA}\\
    \hat Q(x)&=&\frac{i}{2}\int_{-\infty}^\infty dk \sqrt{\frac{\hbar vc|k|}{\pi}}\left(e^{ikx}a_k^\dagger-e^{-ikx}a_k\right)\! .\label{QToA}
\end{eqnarray}
Inserting Eqs.~(\ref{PhiToA}) and (\ref{QToA}) into Eq.~(\ref{HermDisp}), we obtain
\begin{equation} \label{kintermed}
   |\Psi_0\rangle=\exp
   \scalemath{1.0}
   {\left[\frac{1}{2\sqrt{\pi\hbar}}\int\!\! dx\int\!\! dk\ 
   \alpha(k,x)e^{ikx}a_k^\dagger+{\mbox{\em h.c.}}\right]}|0\rangle,
 \end{equation}
 with the shorthand
\begin{equation}
    \alpha(k,x) \equiv \sqrt{cv|k|} \varphi(x)+\frac{iq(x)}{\sqrt{cv|k|}}.
\end{equation}
Note here and in the following, if we give no integration limits, they may be understood to be from $-\infty$ to $\infty$.
With our rewrite of $\Psi$, the creation-operator part of the displacement operator has been isolated here in the first term. We then get the operator that we want by inserting Eq.~(\ref{EigenCreate}) :
\begin{eqnarray}
    \beta b^\dagger&=&\frac{1}{4\pi\hbar}\int dx\int dy\int dk\ e^{ik(x-y)}\\ &&\left[\vphantom{\frac{1}{cv|k|}}\left(\vphantom{\frac{1}{cv|k|}}cv|k|\varphi(x)+iq(x)\right)\hat\Phi(y)\right.\nonumber\\+&&\left.\,\,\,\left(\frac{1}{cv|k|}q(x)-i\varphi(x)\right)\hat Q(y)\right].\nonumber
\end{eqnarray}
The antihermitian parts of this expression work out easily using $\int dk\ e^{ik(x-y)}=2\pi\delta(x-y)$ and confirm the results of Eqs.~(\ref{HermPart}). The two hermitian contributions involve more complicated integrals because of the $|k|$ factors \footnote{It is in carrying out these integrals with the $|k|$ factors without approximation that the present analysis departs from the usual optical approximation of quantum optics; see Sec. 8.1.5 of \cite{gardiner2004quantum}}. These can be worked out as follows: 
\begin{eqnarray}
    \int dx dk|k|\varphi(x)&&\!\!\!\!\!\!\!e^{ik(x-y)}=\int dx dk\, {\mbox{\rm sgn}}(k)\!\cdot\! k\varphi(x)e^{ik(x-y)}\nonumber\\
    \stackrel{\tiny k\rightarrow -k}{=}&&-i\int\! dx dk\,{\mbox{\rm sgn}}(k)
    [\partial_x\varphi(x)]e^{ik(y-x)}
    \nonumber\\
    =&&-i\int dk\, {\mbox{\rm sgn}}(k)e^{iky}\mathcal{F}_k[\partial_x\varphi(x)]
 \nonumber\\
 =&&2\pi\mathcal{F}_y^{-1}\left[-i\,{\mbox{\rm sgn}}(k)\mathcal{F}_k[\partial_x\varphi(x)]\right]
 \nonumber\\
 =&&2\pi\mathcal{H}_y[\partial_x\varphi(x)].\label{bigint1}
\end{eqnarray}
\begin{eqnarray}
    \int dx dk&&\!\!\!\!\!\!\!\!\!\frac{q(x)}{|k|}e^{ik(x-y)}=\int dx dk \frac{q(x)}{k}\,{\mbox{\rm sgn}}(k)e^{ik(x-y)}\nonumber\\
    &=&-i\int dx dk \left(\int_{-\infty}^x dsq(s)\right)\,{\mbox{\rm sgn}}(k)e^{ik(x-y)}\nonumber\\
 &=&i\int dk\, {\mbox{\rm sgn}}(k)e^{iky}\mathcal{F}_{k}\left[\int_{-\infty}^xq(s)ds\right]\nonumber\\
 &=&-2\pi\mathcal{H}_y\left[\int_{-\infty}^xq(s)ds\right].\label{bigint2}
\end{eqnarray}
Here we have used $\mathcal{F}_{k}[f(x)]=\int dx\ e^{-ikx}f(x)$, the Fourier transform, and its inverse $\mathcal{F}_{y}^{-1}[f(k)]=1/2\pi\int dk\ e^{iky}f(k)$, and the Hilbert transform
\begin{equation}
    \mathcal{H}_{y}[f(x)]=\frac{1}{\pi}P\int_{-\infty}^\infty dx\frac{f(x)}{x-y}.
\end{equation}
Properties of this transform, and in particular its relation to the Fourier transform as used in Eqs.~(\ref{bigint1},\ref{bigint2}), can be found in Chap.~15 of \cite{Handbook99}. Note also that Eqs.~(\ref{bigint1},\ref{bigint2}) have used integrations by parts, in particular Eq.~(\ref{bigint2}) uses it in the form
\begin{eqnarray}
    \int_{-\infty}^\infty dx \partial_x\left[\left(\int_{-\infty}^x ds q(s)\right)e^{ikx}\right]&&\\
    = \int_{-\infty}^\infty dx q(x)e^{ikx}+ik 
     \int_{-\infty}^\infty &dx& \left(\int_{-\infty}^x ds q(s)\right)e^{ikx}.\nonumber
\end{eqnarray}

\begin{figure*}[ht]
    \centering
    \includegraphics[width = 0.8\textwidth]{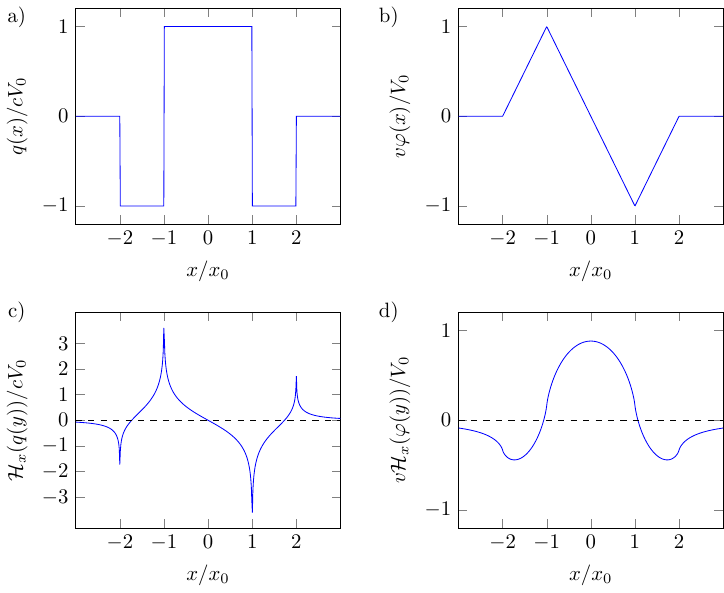}
    \caption{Contributions to the photonic representation Eqs.~(\ref{HermDisp},\ref{creation}) of a simple transmission-line pulse. $x_0$ is an arbitrary length scale; the waveforms shown are independent of $x_0$. a) Assumed charge density (or equivalently voltage) form of the pulse. b) Flux field (or also current field) for which the pulse of part a) is a right-mover. c) Hilbert transform of part a). Note that, consistent with the Paley-Wiener theorem \cite{PaleyWiener}, this function is non-zero outside the support of $V(x)$, and is in fact nonzero for all $x$. The curve features logarithmic divergences at $x/x_0=\pm 1,\pm 2$. d) Hilbert transform of $\varphi(x)$. Referring to Eq.~(\ref{creation}), if the imaginary part of $\theta_q(x)$ is taken to be proportional to the waveform of part a), then part d) is proportional to its real part, part b) is proportional to the imaginary part of $\theta_\varphi$ (assuming a right-mover), and part c) is proportional to its real part. 
    }
    \label{Fig:TrialFunctions}
\end{figure*}
In using this (after dividing by $ik$), we require the left-hand side to be zero, which imposes the nontrivial condition
\begin{equation}\label{con1}
  \int_{-\infty}^\infty ds q(s)=0  
\end{equation}
on AWG pulses that can be studied with our analysis. The other integration by parts used in Eq.~(\ref{bigint1}) leads to the additional condition (it is here that we fix the integration constant of Eq.~(\ref{rel1}))
\begin{equation}\label{con2}
    \varphi(x=\pm\infty)=0.
\end{equation}
We simplify conditions (\ref{con1},\ref{con2}) later for the case of right-traveling AWG pulse waveforms. Note that these conditions are not merely a formality: if they are not satisfied the photonic representation of the pulse, Eq.~(\ref{OneModeDisp}), does not exist. We interpret this to mean that, if Eqs.~(\ref{con1},\ref{con2}) are not satified, then $\beta^2$ (the mean photon number) diverges. 

When our photonic representation exists, we now have a complete solution for the operator in Eq.~(\ref{creation}); the coefficient functions are
\begin{eqnarray}
  \theta_q(x)&=& \mathcal{H}_{x}[cv\partial_y\varphi(y)]+iq(x),\label{fullth1}\\
  \theta_\varphi(x)&=& \mathcal{H}_{x}\left[\frac{1}{cv}\int_{-\infty}^y dsq(s)\right]+i\varphi(x).\label{fullth2}
\end{eqnarray}
We can also get a general expression for $\beta^2$, the expected number of photons in the pulse. We impose the condition $[b,b^\dagger]=1$, fixing $\beta$ from the value of the commutator of the two terms in Eq.~(\ref{kintermed}), and use $[a_k,a_k'^\dagger]=\delta(k-k')$. We obtain \footnote{Besides being manifestly positive, two other properties of $\beta^2$ that we confirm are that $\beta^2$ is a constant under time evolution, and that $\beta^2$ is additive, that is, it is the sum of a left- and right-moving contribution.}
\begin{equation}\label{generalbeta}
\scalemath{1.00}{\beta^2=\frac{1}{4\pi\hbar}\int dk\left(cv|k|\cdot\left\vert\mathcal{F}_k[\varphi(x)]
\,\right\vert^2+\frac{\left\vert\mathcal{F}_k[q(x)]\,\right\vert^2}{cv|k|}\right)}.
\end{equation}

\section{Right-Moving Pulses}

Eqs.~(\ref{fullth1},\ref{fullth2},\ref{generalbeta}) can be considered a final result, and we will examine the surprising consequences of these formulas in several examples. But for these examples, we will impose the further condition discussed earlier that the pulse is right-moving. This results in an interesting simplification of the expressions for $\theta_q$, $\theta_\varphi$ and $\beta$.

From Eq.~(\ref{rightmove}), the right-moving condition is 
\begin{equation}
    V(x)=Z_0 I(x)
    \label{VI}
\end{equation}
Using Eqs.~(\ref{rel1},\ref{rel2}), we can then determine both our functions $q(x)$ and $\varphi(x)$ solely from $V(x)$:
\begin{equation}
    q(x)=cV(x)\label{rm1}
\end{equation}
\begin{eqnarray}
    \frac{d\varphi(x)}{dx}&=&-\frac{1}{vc}q(x),\\
    \Rightarrow\varphi(x)&=&-\frac{1}{vc}\int_{-\infty}^x dsq(s),\\
    \varphi(x)&=&-\frac{1}{v}\int_{-\infty}^x dsV(s).\label{rm2}
\end{eqnarray}
Note that these right-mover conditions also reduce the two conditions Eqs.~(\ref{con1},\ref{con2}) for the validity of the photonic representation of the state Eq.~(\ref{OneModeDisp}) to the single condition
\begin{equation}
    \int_{-\infty}^\infty dsV(s)=0.\label{oneVPC}
\end{equation}
Since an AWG can readily create a pulse with a nonzero average $V$, this is a significant restriction.

Applying the right-moving conditions Eqs.~(\ref{rm1},\ref{rm2}), our coefficient functions $\theta_q$ and $\theta_\varphi$, and the displacement amplitude $\beta$, take the simplified form
\begin{eqnarray}
  \theta_q(x)&=&-\mathcal{H}_{x}[cV(y)]+icV(x),\label{fullth1rm}\\
  \theta_\varphi(x)&=&\mathcal{H}_{x}\left[\frac{1}{v}\int_{-\infty}^y dsV(s)\right]-\frac{i}{v}\int_{-\infty}^x dsV(s),\label{fullth2rm}\\
  \beta^2&=&\frac{1}{2\pi\hbar}\frac{c}{v}\int\frac{dk}{|k|}\left|\vphantom{\frac{1}{\sqrt{|k|}}}\mathcal{F}_k[V(x)]\right|^2.\label{fullth3rm}
\end{eqnarray}
We get an alternative expression for $\beta^2$ by doing the $k$ integral in (\ref{fullth3rm}). Using also the bipolar condition on $V(x)$, Eq.~(\ref{oneVPC}), we get
\begin{equation}
    \beta^2=\frac{1}{\pi\hbar}\frac{c}{v}\,\!\int\!\!\int dxdyV(x)V(y)\ln(|x-y|).
\end{equation}
This equation has the appealing form as a quadratic integral expression in $V(x)$ with a translationally invariant kernel, and is quite practical for explicit calculations. Bipolarity also makes this scale invariant, i.e., independent of the units in which $x$ and $y$ are measured. A similar simplification of the more general expression for $\beta^2$, Eq.~(\ref{generalbeta}), can be attempted, but in this case the translationally-invariant kernel multiplying $\varphi(x)\varphi(y)$ is highly singular and not practical for calculations.

Combinations of the form in $\theta_q$ and $\theta_\varphi$ in Eqs.~(\ref{fullth1rm},\ref{fullth2rm}) have a special name in signal processing theory (Chap.~15, \cite{Handbook99}) -- they are called {\em analytic signals} (see also discussions in \cite{bialynicki1994wave,Saari06,ROY2016740}). In particular, the function $\theta_q$ is $(ic)$ times the analytic signal of the waveform $V(x)$, while the function $\theta_\varphi$ is $(-i/v)$ times the analytic signal of the once-integrated waveform $\int_{-\infty}^x dsV(s)$. 



\section{Examples}

We only look at toy examples here, realistic pulses could be analyzed with the aid of modern algorithms for computing the Hilbert transform \cite{Bilato2014}. We begin with a simple, bipolar, right moving square voltage pulse (Fig.~\ref{Fig:TrialFunctions}a). We must have its integral (Fig.~\ref{Fig:TrialFunctions}b), and their Hilbert transforms (c and d). Note a crucial property of these Hilbert transforms, which is that they {\em extend beyond the support of the voltage pulse}. In fact, they are nonzero all the way to infinity. This is a mandatory property of the Hilbert transform of any bounded-support function, expressed as the Paley-Wiener theorem of signal processing \cite{PaleyWiener}. This property will be key in the considerations of how these photons are measured, discussed in the next section.  

For the simple example voltage pulse of Fig.~\ref{Fig:TrialFunctions}a, one gets the result
\begin{equation}
    \beta^2=\frac{12\ln{(\sfrac{27}{16})} x_0^2 V_0^2}{\pi\hbar}\frac{c}{v}
\end{equation}
We note some of the scaling properties of this result. Recall that $\beta^2$ is the expectation value of the photon number of our coherent state. We note:
\begin{equation}
    \langle n\rangle=\beta^2=C\frac{V_0^2x_0^2}{\hbar}\frac{1}{v^2Z_0}=C\frac{1}{\hbar}\cdot\frac{V_0^2 T}{Z_0}\cdot T.\label{neff}
\end{equation}
Here $C$ is some constant and $T=x_0/v$ sets the scale of the transit time of the pulse. The last part of Eq.~(\ref{neff}) shows that this expression can be viewed as being in the familiar form $E/\hbar\omega$, where $E$, the energy of the pulse, goes as $V_0^2T/Z_0$, and $\omega$ is identified with $T^{-1}$.

\begin{figure}[ht]
    \centering
    \includegraphics[width = \columnwidth]{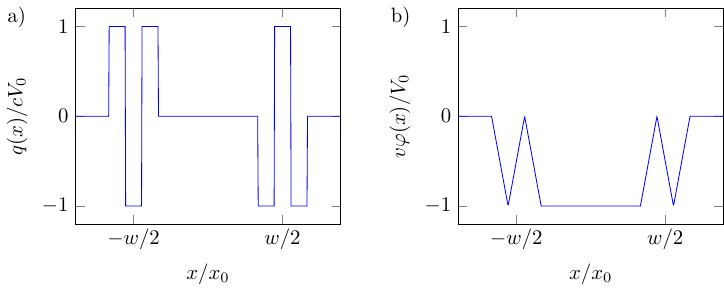}
    \caption{a) Train of two short voltage pulses separated by normalized distance $w$. Unlike in Fig.~\ref{Fig:TrialFunctions}, each pulse individually has 3 intervals of equal length $x_0$ at which the charge density expectation value is $+cV_0$ or $-cV_0$. Therefore the short pulses do not integrate to zero separately and they must be considered together. The total mean photon number scales like $\log w$. b) $\varphi(x)$, or equivalently current profile, which is seen to be nonzero between the two pulses.}
    \label{Fig:DoublePulse}
\end{figure}

For the next example, we introduce a train of two pulses similar to 
Fig.~\ref{Fig:TrialFunctions}; this new double pulse is shown in Fig.~\ref{Fig:DoublePulse}. But note that in detail Figs.~\ref{Fig:TrialFunctions} and \ref{Fig:DoublePulse} are different, and, crucially, each of the two pulses in  Fig.~\ref{Fig:DoublePulse} does not individually integrate to zero. But, because the second pulse is inverted compared to the first, the integral over both is zero. It is clear from above that these two pulses must be treated as one in analyzing their photonic content. This is also clear from Fig.~\ref{Fig:DoublePulse}b, where we see that $\varphi(x)$ is nonzero in the whole interval between the two pulses.

\begin{figure*}[ht]
    \centering
     \includegraphics[width = 18cm]{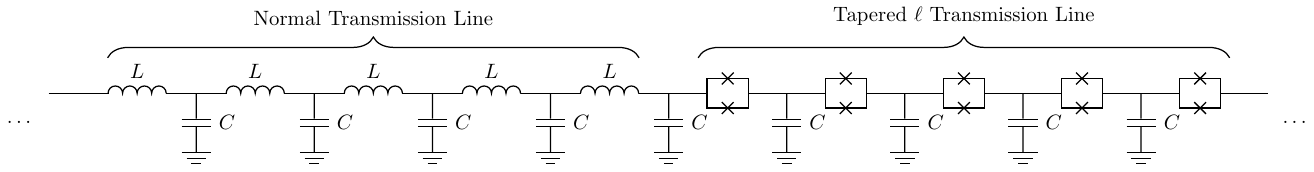}
    \caption{Part of gedanken apparatus for trasfer or detection of pulse photons. One aspect of an optimized apparatus can be to bring the pulse, without other disturbance, to a much slower velocity. It is suggested that this may be done by connecting the normal transmission line to a metamaterial (SQUID-based) transmission line, in which the velocity $v$ is smoothly ramped towards zero by the gradual tapering of the inductance per unit lenth $\ell$, such that $1/\ell\rightarrow 0$.}
    \label{Fig:TransmissionLine}
\end{figure*}

For the signal of Fig.~\ref{Fig:DoublePulse}, we will only discuss the result for the mean photon number. The integration Eq.~(\ref{fullth3rm}) can be done for arbitrary separation between the two pulses $wx_0$. The exact result is lengthy, but asymptotically gives
\begin{equation}
    \beta^2\propto\ln{w}.
\end{equation}

\section{Detection of photons in transmission-line pulse}

We do not have realistic experiments to propose that would measure photons in the pulses that we have discussed above. But we can, at the level of gedanken experiments, indicate strategies that could be usefully pursued in the development of some experimental approaches.

We can say that we must ``measure $b$" (Eq.~(\ref{creation})) in a photon counting experiment, or ``measure $b^\dagger b$" in a quantum non-demolition experiment. We need to be more specific than this, but one point to note is that there will be a difficulty because neither $b$ nor $b^\dagger b$ commutes with the transmission-line Hamiltonian Eq.~(\ref{hamtrans}) -- our photons, and indeed photons generally, do not have the attribute of having a definite frequency \cite{ROY2016740}. Thus, one might imagine that a helpful step in the measurement process would be to turn off the transmission line Hamiltonian. This is at least partially accomplished if the techniques of {\em slow light} and {\em stopped light} are applied to our microwave pulse. We refer to techniques that were developed some time ago for optical radiation \cite{hau2001frozen}, and are recently considered in the far-infrared regime \cite{zhao2019dual}.

A version of this for superconducting transmission lines in the microwave band could be using tunable, metamaterial transmission lines \cite{castellanos2008amplification}. With reference to Fig.~\ref{Fig:TransmissionLine}, a normal transmission line can transition into one with a gradually larger $\ell(x)$, by means of the flux biasing of the metamaterial, which here is simply a one-dimensional array of SQUIDs, whose effective inductance is varied by an external flux. $1/\ell$ can even be made to vanish \cite{castellanos2008amplification}. We suggest that with a suitable tapering to slow a pulse adiabatically, combined with a switching, at the right moment, to the condition $1/\ell=0$, one can bring our pulse to a halt without essentially changing its quantum state.

The adiabatic slowing, and sudden freezing, which we have just described should not change the shape of the pulse, although it will be compressed spatially. This could be a plus, since, for example, a 3ns pulse will, on a conventional transmission line, extend over a large fraction of a meter. Compression by, say, a factor of 1000 would make the region to be measured a more convenient millimeter-scale size.

But can the measurement take place just where the voltage pulse is nonzero? Yes, but only a clearly sub-optimal measurement. We now consider some of the general properties of the optimal measuring instrument, used to measure the frozen pulse. We use the phrase ``measuring instrument" in the way meant in the tripartite measurement theory of von Neumann, see 
Sec.~VI.1 of \cite{von2018mathematical}, where it is called ``part II" of the setup. (Part I is the system to be measured, and part III is the ``observer". Von Neumann has an interesting discussion of the non-uniqueness of the boundaries between these three subsystems.)

\begin{figure*}[ht]
    \centering
    \includegraphics[width = 17cm]{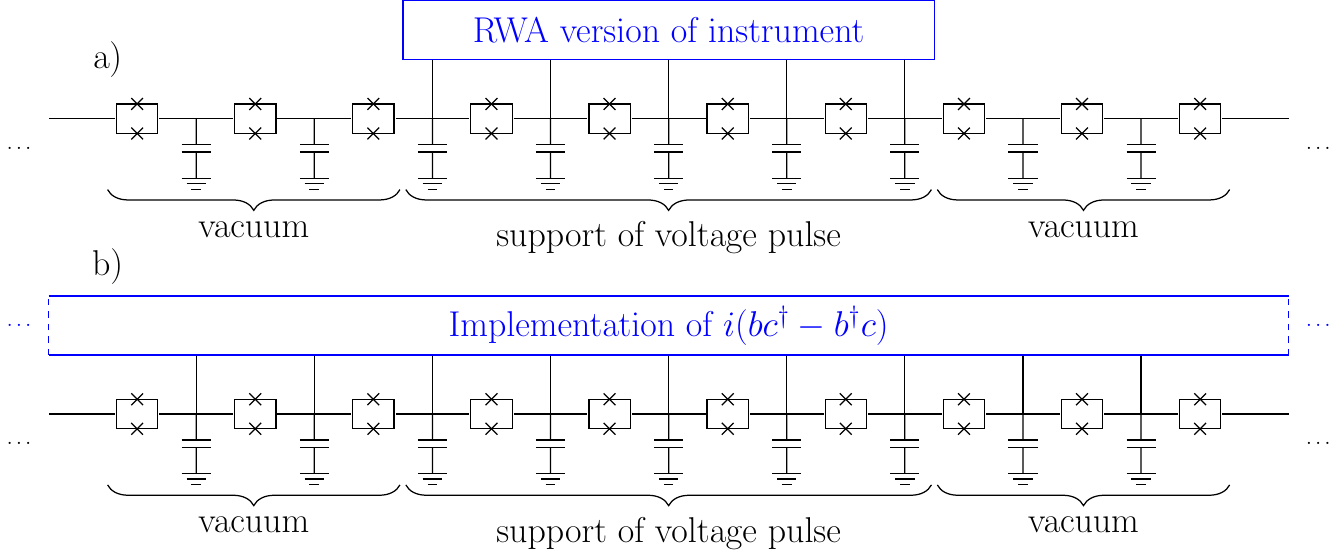}
    \caption{General features of the measuring instrument. In the stopped-pulse scenario, the instrument must interact with the transmission line over an extended distance. a) To implement interaction $i(b-b^\dagger)(c+c^\dagger)$, it is sufficient for the interaction to extend just over the support of the voltage pulse. But this is only an RWA (rotating wave approximation) to the optimal photon-counting coupling $i(bc^\dagger-b^\dagger c)$. But the RWA is not a good approximation when pulses do not have a well-defined frequency. b) The implementation of the actual optimal interaction $i(bc^\dagger-b^\dagger c)$ requires interaction with the transmission like extending far into the ``vacuum" region, where the voltage expectation value is zero.}
    \label{Fig:Coupler}
\end{figure*}

Since apparatus II is only intended to record an integer (a photon count), it suffices to single out one bosonic degree of freedom internal to the apparatus, whose operators we will call $c$ and $c^\dagger$. With this, we can first focus on a common form of coupling that would be chosen for a photon counter:
\begin{equation}
    H_{\text{int}}\propto i(b-b^\dagger)(c+c^\dagger).
    \label{measRWA}
\end{equation}
It is clear that this interaction could be implemented by an instrument that physically couples only to the section of transmission line containing the pulse (Fig.~\ref{Fig:Coupler}a), since we note from Eqs.~(\ref{HermDisp},\ref{OneModeDisp}) that $i(b-b^\dagger)$ is proportional to the field operator
\begin{equation}
    q(x)\hat\Phi(x)-\varphi(x)\hat Q(x)=cV(x)\hat\Phi(x)+\frac{\hat Q(x)}{v}\int_{-\infty}^x  dsV(s),
\end{equation}
which is zero in the vacuum region of the transmission line.
Note that the quadrature chosen for the system part of $H_{\text{int}}$ in (\ref{measRWA}) matters for this conclusion.

But note furthermore that there is no reasonable RWA (rotating wave approximation) that justifies the replacement of (\ref{measRWA}) by the desired interaction,
\begin{equation}
H_{\text{int}}'\propto i(bc^\dagger-b^\dagger c),
\label{optHint}
\end{equation} 
which would describe the desired transfer of quanta from system to instrument for detection. If an apparatus II for the optimal counting interaction Eq.~(\ref{optHint}) can be built, it must have the feature, as shown in Fig.~\ref{Fig:Coupler}b, that it interacts with the transmission line also in the vacuum region. Actually, according to the Paley-Wiener theorem, this interaction would actually have to extend to infinity. While we have not investigated the question quantitatively, but we expect that a very good approximation to the optimal counting measurement would be achieved by a finite interaction region, so long as it extends well into the vacuum.

How can it possibly be useful to ``measure the vacuum" when one is trying to count photons in a pulse? We would claim that this is due to the famous observation of Summers and Werner that the bosonic quantum vacuum is entangled \cite{summers1985vacuum}\footnote{See work of Reznik and coworkers \cite{botero2004spatial} for more recent studies of this phenomenon for the 1D bosonic vacuum as considered in the present paper.}, applying more generally to   the vacua of other quantum field theories, as in the work of Reeh and Schlieder \cite{ReehSchlieder1961,Haag}. Consequently, we view the optimized measurement discussed here as an instance of entanglement assisted measurement \cite{degen2017quantum,huelga1997improvement}, first seen in the concept of superdense coding in quantum communication theory \cite{Bennett1992}. Note that the inclusion of vacuum in the optimal measurement clearly also extends to the QND version of the optimal measurement, which would be governed by the interaction Hamiltonian
\begin{equation}
H_{\text{int}}^{\mbox{\tiny{\em QND}}}\propto b^\dagger bc^\dagger c.
\end{equation}

The results obtained here gives an interesting perspective on future attempts to improve on quantum cyrptography with dim coherent states. It may be important to explore the shortest possible pulses, in order to improve on key exchange rates in the crypto protocol. Danger arises if the users settle on a non-optimal measurement scheme, e.g., one which is sensitive only in a band of frequencies. The present work emphasizes that a photon has no definite frequency, and that pulses that appear very dim ($\langle n\rangle\ll1$ from the point of view of traditional detectors are (see Fig.~\ref{Fig:DoublePulse}) actually very bright, with an arbitrarily large $\beta^2$. Eve, in possession of a QND detector of the sort described above, would easily break a key distribution system which Alice and Bob, with wide-band frequency detectors, feel mitakenly to be secure.

We regret that our story is presently incomplete, in the sense that while our part-II detectors above are unquestionably possible in principle, we do not know presently how they would actually be constructed with the tools of circuit-QED \cite{blais2021circuit}. We must admit that we do not have as complete a view of our photon detectors as those constructed in the recent work of M\o{}lmer and coworkers \cite{PhysRevA.107.013705,PhysRevA.107.013706}. It will certainly be a good challenge for this remarkably powerful toolkit to attempt this construction in a later work.





\section{Acknowledgements}

D.D.V. acknowledges discussions with Guido Burkard concerning Ref.~\cite{virally2019unidimensional}, and project work of Sashank Kaushik Sridhar on preliminary aspects of the present work \cite{sridhar2021active}. We acknowledge support from the Deutsche Forschungsgemeinschaft (DFG) 
under Germany's Excellence Strategy Cluster of Excellence Matter and Light for Quantum Computing (ML4Q) EXC 2004/1 390534769.

\bibliography{references.bib}

\end{document}